\documentclass{jps-cp}
\usepackage{txfonts}
\usepackage{url}
\usepackage{amsmath}
\usepackage{afterpage}
\usepackage{url}
\newcommand\newblock{\hskip .11em plus .33em minus .07em}
\newcommand\figref[1]{Fig.~\ref{#1}}
\newcommand\tabref[1]{Table~\ref{#1}}

\title{XRP Network and Proposal of Flow Index}
\author{Hideaki Aoyama}
\inst{Graduate School of Advanced Integrated Studies in Human Survivability, Kyoto University, Kyoto 606-8306, Japan\\
RIKEN iTHEMS, Wako, Saitama 351-0198, Japan\\
Research Institute of Economy, Trade and Industry (RIETI), Tokyo 100-0013, Japan}
\email{hideaki.aoyama@gmail.com}
\recdate{May 17, 2021}

\newcommand\ihh{Inverse Herfindahl-Hirschman Index}

\abst{
XRP is a modern crypto-asset (crypto-currency) developed by Ripple Labs, which has been increasing its financial presence. 
We study its transaction history available as ledger data.  
An analysis of its basic statistics, correlations, and network properties are presented.
Motivated by the behavior of some nodes with histories of large transactions, we propose a new index: the ``Flow Index.'' The Flow Index is a pair of indices suitable for characterizing transaction frequencies as a source and destination of a node.
Using this Flow Index, we study the global structure of the XRP network and construct bow-tie/walnut structure.\\[5pt]
{\sl Presented at the``Blockchain in Kyoto" (February 17--18, 2021), and will appear in the JPS Conference Proceedings.}}

\kword{Power distribution, Ripple Labs., Ledger data, bow-tie/walnut structure}
\begin{document}
\maketitle

\section{Introduction}

The world of crypto-assets is dynamic and complex (we use the term ``crypto-asset'' instead of ``crypto-currency'' throughout this paper.
See \cite{crypto}).
Its presence in the financial market has steadily increased since its inception in early 2013.
Understanding the nature of this world is important.

Ever since its inception, all transaction data (except for a few, which we will elaborate in the next section) are stored and available through various media, providing researchers with ample opportunity to study the intriguing properties, similar to Bitcoin.
\cite{islam2019analyzing,islam2020unfolding,fujiwara2020hodge}.

Traditional monetary transactions through financial institutions, such as banks, are crucial for analyzing and understanding inter-firm and firm-household relationships. However, the availability of such data is quite limited because of privacy concerns, except for a few rare cases \cite{yoshi2021money}.

In this study, we present a basic analysis of the XRP world observed through ledger data, which is the transaction record comprising the amount of XRP, source account, destination account, and the day and time (in coordinated universal time (UTC)) of transactions.
(For a reasonable and readable introduction to XRP, see \cite{XRPintro}.)
Accounts are just 33 letter-long codes such as ``rfceigRxmgAjWR6LH1L7YsooWKMqM5Pr6,'' and no other information on the owner (name, address, etc.) are available.
Although this makes interpreting the results of analysis  rather difficult, because of its importance and data availability, it remains a worthwhile endeavor.

In Section 2, we describe our ledger data and its basic properties,
including the distribution of the number of transactions, properties of the time series with the day-of-the-week analysis, and correlations.
Section 3 is devoted to analyzing the XRP network of the transactions, wherein nodes are accounts and edges are transactions.
Section 4 describes the new proposal of the {\it Modified \ihh} and {\it Flow Index}.
Section 5 is devoted to studying the global structure of the 
XRP network, similar to bow-tie/walnut decomposition, using the Flow Index.
Section 6 offers discussion and conclusion.

\section{Data and its Basic Statistics}
The ledger data we analyze are for 2,463 days of 1/2/2013--9/30/2019.
The first 32,570 ledgers were lost because of ``a mishap in 2012'' \cite{mishap}.

\subsection{Data Selection}
\begin{figure}[t]
    \centering
    \includegraphics[width=0.6\textwidth]{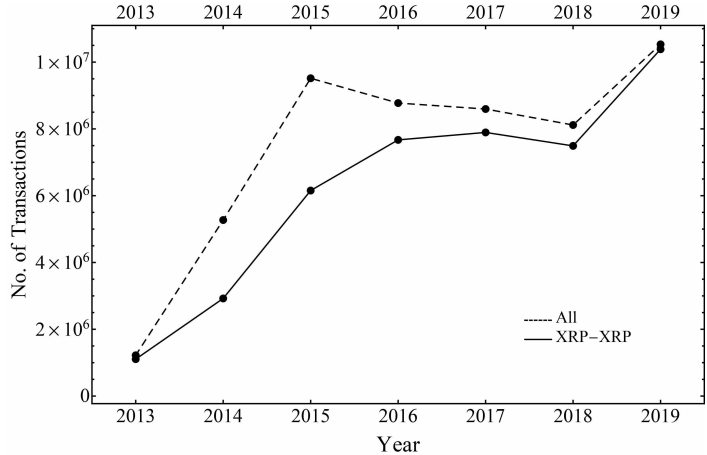}
    \caption{Annual number of transactions. The points connected with dashed lines denote all transactions, whereas those with solid lines denote XRP--XRP.}
    \label{fig:nt}
\end{figure}

\begin{figure}[t]
    \centering
    \includegraphics[width=0.5\textwidth]{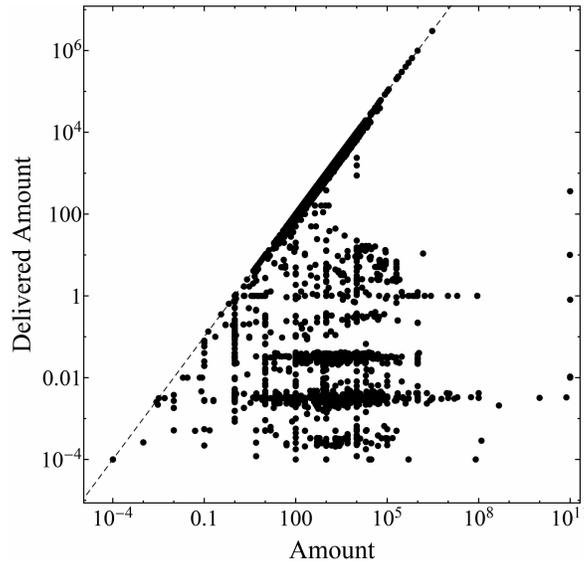}
    \caption{Distribution of ``Amount" and ``Delivered amount'' of the XRP--XRP transactions whose ``Amount" is not equal to ``Delivered Amount.'' The dashed diagonal has a gradient equal to one wherein the amount is equal to the delivered amount.}
    \label{fig:del}
\end{figure}

From this data set, we extract data with the following criteria.
\begin{enumerate}
\item 
The ledger contains transactions between 1,525 currencies and crypto-assets.
Most of these are from XRP to XRP; however, some of them are from XRP to others, others to XRP, and others to others.
We provide a yearly breakdown for those with XRP--XRP, and the rest is provided in Fig.\ref{fig:nt}.
We use XRP--XRP transactions only.
\item
The data set contains ``partial payments'' \cite{pp}, wherein the actual transferred amount is different from the transaction amount because of the payment of transfer fees.
In reality, they are rare:
0.018\% of all of the XRP--XRP transactions, with the minimum of 0\% in 2013 and a maximum of 0.041\% in 2019.
The distribution of ``Amount'' and ``Delivered amount'' and the delivered amount is provided in \figref{fig:del},
wherein we observe some patterns of quantization of the delivered amount and the proportionality between the amount and the delivered amount.
We drop these ``partial payment'' transactions from the following analysis.
\end{enumerate}

\begin{table}[t]
    \centering
    \begin{tabular}{c|c|r}
Source & Destination & No.\ of Transactions\\
\hline
XRP&	XRP&	43,624,956 \\
CCK&	CCK&	3,836,798\\
CNY&	CNY	& 933,208\\
EUR&	EUR&	836,061\\
USD&	USD&	667,504\\
SFO&	SFO&	487,965\\
BTC&	BTC&	473,921\\
JPY&	JPY&	207,385\\
ETH&	ETH&	168,908\\
GWD&	GWD&	146,054
    \end{tabular}
    \caption{Top 10 transaction currencies.}
    \label{tab:t1}
\end{table}

After filtering, we arrive at the data of the sizes listed in Table \ref{tab:ydata}.

\begin{table}[tb]
    \centering
    \begin{tabular}{c|r|r|r|r|r}
       Year & \# Transactions&\hspace{17pt}\# Source & \# Destination & \# All nodes & XRP ($/10^{12}$)\\
       \hline
2013&	1,104,589&22,410&	548,38&	54864&	0.2031\\
2014&	2,897,049&35,477&	130,658&	131,525&	0.1368\\
2015&	6,147,511&28,362&	56,724&	64,310&	0.0625\\
2016&	7,652,661&31,370&	68,167&	75,230&	0.2939\\
2017&	7,883,617&297,558&	666,749&	693,787&	0.2109\\
2018&	7,483,668&424,695&	714,184&	826,622&	0.1310\\
2019&	10,379,060&378,700&	434,675&	535,724&	0.1266\\
\hline
All years&	43,548,155&1,287,516&	1,810,387&	1,810,676&	1.1651\\
    \end{tabular}
    \caption{Number of transactions, nodes (source, destination, and either of them) and the amount of traded XRP.}
    \label{tab:ydata}
\end{table}

\subsection{Data Distributions}
\begin{figure}[t]
    \centering
    \includegraphics[width=0.6\textwidth]{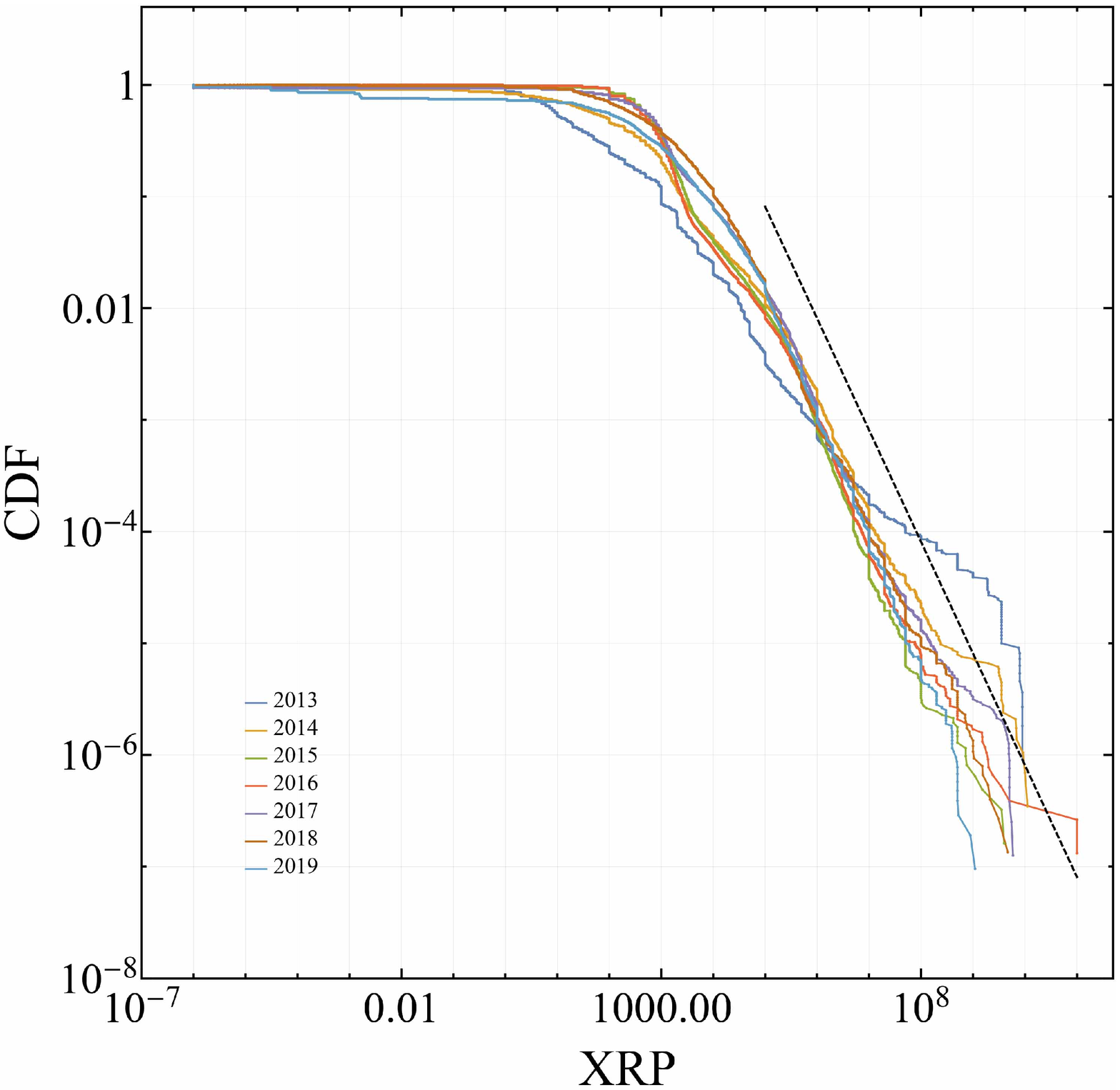}
    \caption{Annual Cumulative Distribution Function (CDF) of the amount of XRP transactions.}
    \label{fig:XRPCDF}
\end{figure}

The cumulative distribution function (CDF) of the annual XRP transaction is plotted in \figref{fig:XRPCDF}, wherein  the dashed straight line has a gradient of $=-1$.
The data for XRP $>10^5$ fit are appropriate to this line, except for the first year of 2013.
This means that the XRP distribution has a power-decaying fat tail,
XRP$^{-1}$ in CDF and XRP$^{-2}$ in the Probability Distribution Function (eps).
The absolute value of this exponent of CDF tail is called the ``Pareto index.".
The current Pareto index 1 is known to be a phase transition point between an ``Oligopoly'' phase and a ``Pseudo Equality'' phase
\cite{aoyama2010econophysics,aoyama2017macro}:
In case the fat tail is thinner and the Pareto index is larger than 1,
the share of those with higher ranks (top, largest, second largest, and so forth) have zero shares when an infinite number of entries are present.
In contrast, the top-ranking ones have finite shares, even when an infinite number of entries exists.

The top 10 transactions (in amount) are listed in \tabref{tab:toptra}.
As shown in \tabref{tab:ydata}, the total traded amount is $\sim 1.1651\times 10^{12}$ XRP.
Therefore, the top two transactions in this table occupy 17\% of all $\sim45$ million transactions, which is indeed a large share.

\begin{table}[b]
    \centering
    \begin{tabular}{c|rccc}
    No.&XRP&Destination&Data \& Time&Source\\
    \hline
1&	1.0$\times 10^{11}$&	bn0864&	2016-11-07T07:50:20Z&	bn0347\\
2&	1.0$\times 10^{11}$&	bn0864&	2016-11-07T07:51:10Z&	bn0347\\
3&	1.1$\times 10^{10}$&	bn0530&	2014-06-10T21:59:40Z&	bn0598\\
4&	1.0$\times 10^{9}$&	bn0166&	2014-06-10T22:01:50Z&	bn0915\\
5&	9.1$\times 10^{9}$&	bn0151&	2014-06-10T22:07:40Z&	bn0781\\
6&	9.0$\times 10^{9}$&	bn0545&	2013-01-26T22:35:20Z&	bn0103\\
7&	9.0$\times 10^{9}$&	bn0760&	2013-01-26T22:36:00Z&	bn0103\\
8&	9.0$\times 10^{9}$&	bn1010&	2013-01-26T22:39:00Z&	bn0103\\
9&	9.0$\times 10^{9}$&	bn0298&	2013-01-26T22:39:30Z&	bn0103\\
10&	8.0$\times 10^{9}$&	bn0135&	2013-01-26T22:36:40Z&	bn0103
    \end{tabular}
    \caption{Top 10 transactions. ``bnxxx'' is a encoded node name defined by the author.}
    \label{tab:toptra}
\end{table}

This criticality of the Pareto index being equal to one is most easily understood when referring to the case of the size of firms in a country.
Empirical analysis of firms in developed countries (such as Japan, France, Germany, and the United Kingdom) showed that the firm size (number of employees, amount of sales, or income) distribution has a Pareto index very close to one through many years.
This is in contrast to the distribution of personal income whose Pareto index varies around 2 (say, 1.5--2.5), depending on the economic situation of the year.
For firms, business competition drives the Pareto index downward (fatter tail), as big firms attempt to dominate the market. In contrast, various political pressures and measures by the central bank and the ministries against monopoly and oligopoly are active. 
The current author argued that the balancing critical point is at a Pareto index equal to one \cite{aoyama2010econophysics}.
However, a big difference between this argument on firm size and the current XRP transaction exists.
The former is ``stock,'' while the latter (the transaction amount we analyze) is ``flow.''
Moreover, the XRP world is free from central governing organization and there is no measure against monopoly and oligopoly.
Thus, the reason behind the current finding remains a mystery for the  moment.

\subsection{Time Series}
\begin{figure}
    \centering
     \includegraphics[width=1\textwidth]{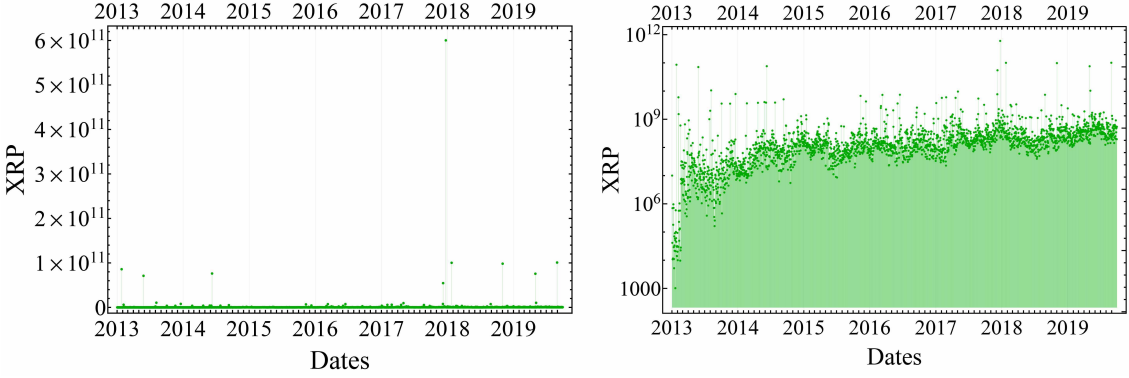}
    \caption{Daily amount of transactions in linear (left) and log scale (right).}
    \label{fig:aud}
\end{figure}

\begin{figure}
    \centering
    \includegraphics[width=1\textwidth]{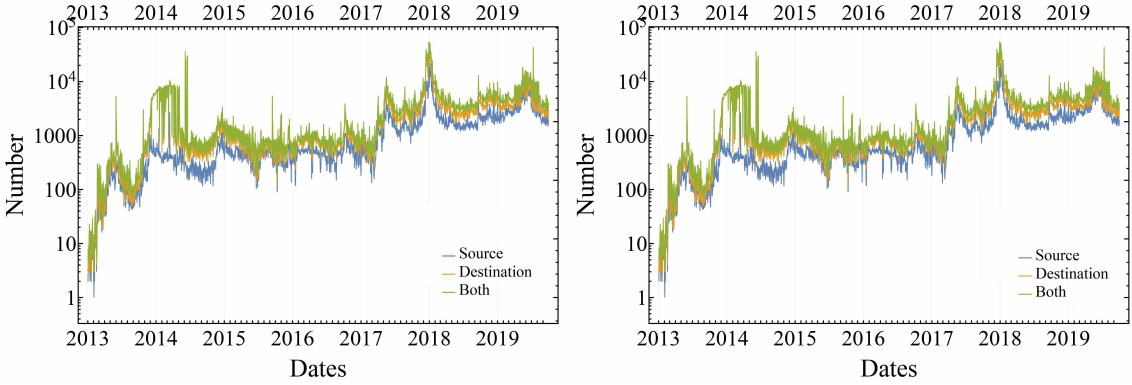}
    \caption{Daily number of users in linear scale (left) and log scale (right).}
    \label{fig:nud}
\end{figure}

\figref{fig:aud} shows the daily amount of transactions.
\figref{fig:nud} shows the daily number of users (blue, orange, and green lines show number of sources, destinations, and either sources or destinations, respectively).
We observe that both the amount and number of users are highly volatile and that most users trade both as a source and destination.

\begin{figure}
    \centering
    \includegraphics[width=0.49\textwidth]{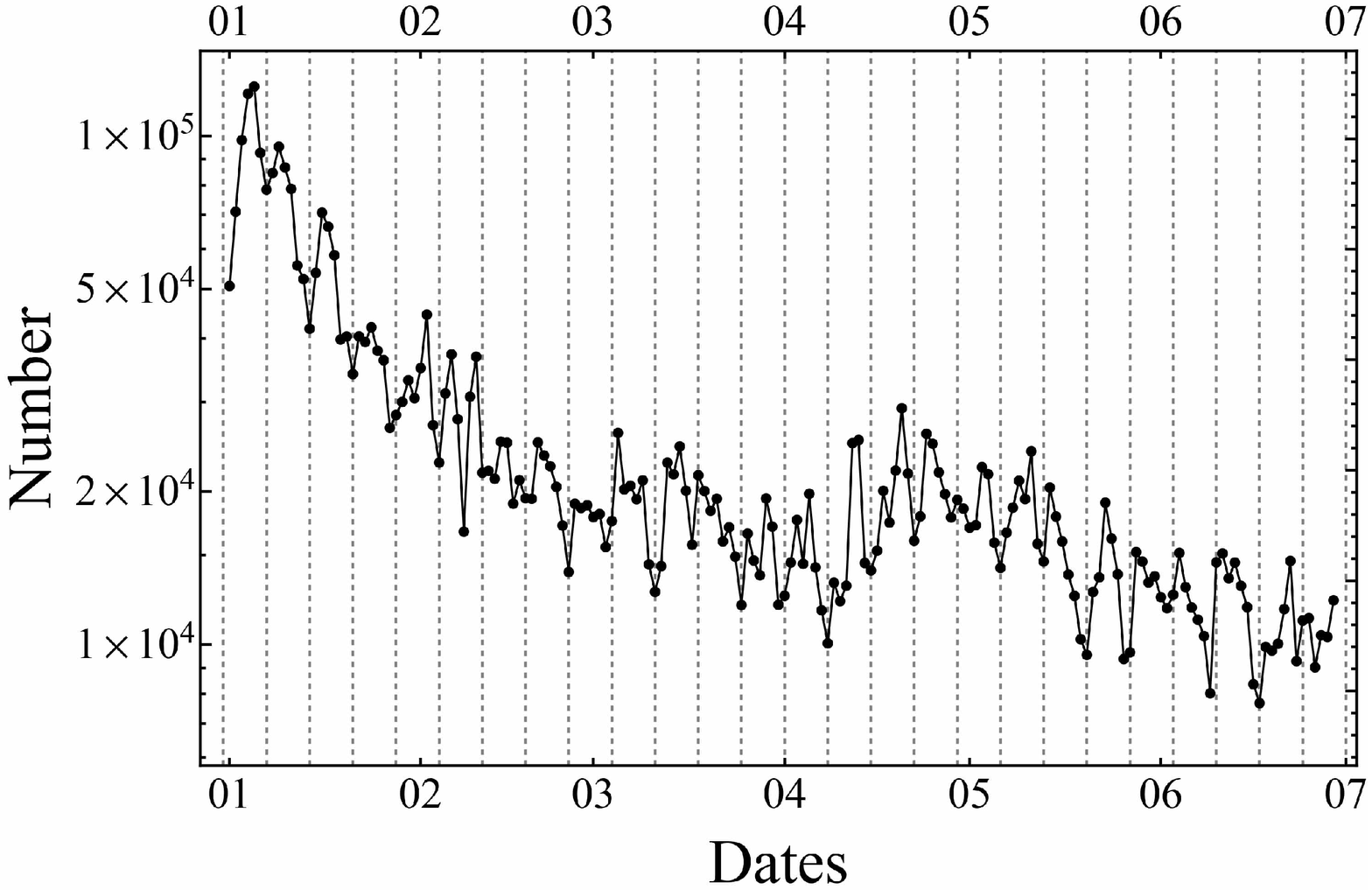}
     \includegraphics[width=0.49\textwidth]{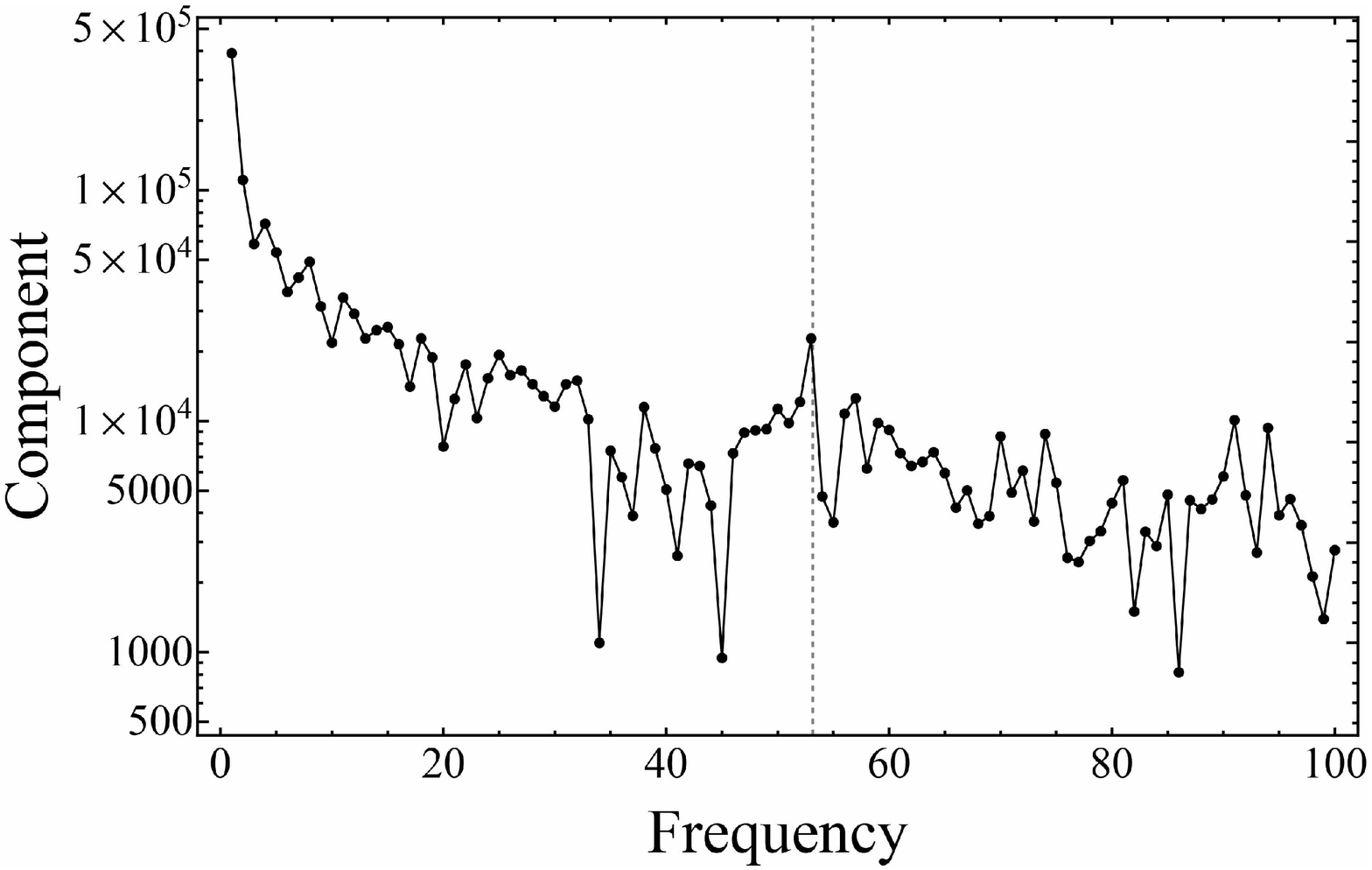}
    \caption{Detail of the daily transaction numbers in 2018.
    The left panel is the daily series, and the right panel is its Fourier decomposition.}
    \label{fig:dday}
\end{figure}

The left panel of \figref{fig:dday} shows the details of the daily number of transactions in 2018, wherein Sundays (in UTC) are shown with a vertical dashed line. 
The reduction of transactions on Saturdays and Sundays is visible to the naked eye. This indicates that most of the nodes are operated by humans.
(Although the time is in UTC and the data cover the entire world, the Saturdays and Sundays in UTC are approximately weekends in most economically active regions, as the Pacific Standard time (PST) in the United States is UTC-7, and Japan Standard Time (JST) is UTC+9.) 
It is clear that the absolute values of the Fourier components in the right panel present a clear peak at a period of seven days. 
The daily number of users also shows reduction on weekends. 
However, the daily total amount of transactions does not show a clear periodicity. This may be because of its high volatility.
This weekly periodicity is weaker in other years.
A similar behavior was found in the analysis of the volume and number of Bitcoin transactions \cite{islam2019analyzing}.

\afterpage{\clearpage}
\subsection{Correlation}
\begin{figure}
    \centering
    \includegraphics[width=0.6\textwidth]{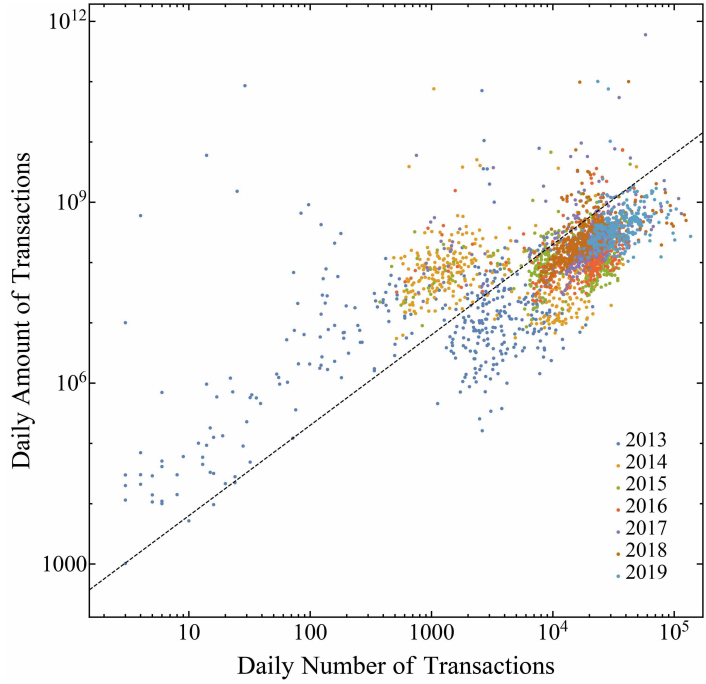}
    \caption{Scatter plot of daily number of users and daily total amount of transactions.}
    \label{fig:nacol}
\end{figure}

There is a correlation that obeys an interesting phenomenological law.
\figref{fig:nacol} shows the correlation between daily number of users and daily total amount of transactions, in which different colors show different years.
The dashed line has gradient = 1.5, which means that
\begin{equation}
    \textrm{[Amount]}\propto \textrm{[Number of users]}^{1.5}.
    \label{prop}
\end{equation}
We observe yearly development toward large numbers of users and a larger amount of transactions roughly along this line. A close examination of the annual data reveals that the distribution split into a group above and another below this line in 2013 and 2014, respectively; however, this converges in later years.
This power law is curious, calling for the modeling of agents in this XRP world:
Since Eq.\eqref{prop} means
\begin{equation}
   \textrm{[Amount per user]}\propto \textrm{[Number of users]}^{0.5},
    \label{prop2}
\end{equation}
This means an interesting characteristic  of the herding behavior in XRP trade: in a day of high activity with a number of users larger than usual, the amount of average transactions increases.
For example, if the number of users becomes 10 times as much, the amount of average transactions becomes  $\sqrt{10}\simeq 3.2$ times as much.

Other correlations between the number of destinations, number of sources, and number of transactions show no such behavior. All three correlations follow linear proportionality (power exponents are close to one).  

\section{XRP Network}
Let us examine the network(s) they form, where nodes are accounts, and edges are transactions.

\begin{figure}[b]
    \centering
    \includegraphics[width=0.49\textwidth]{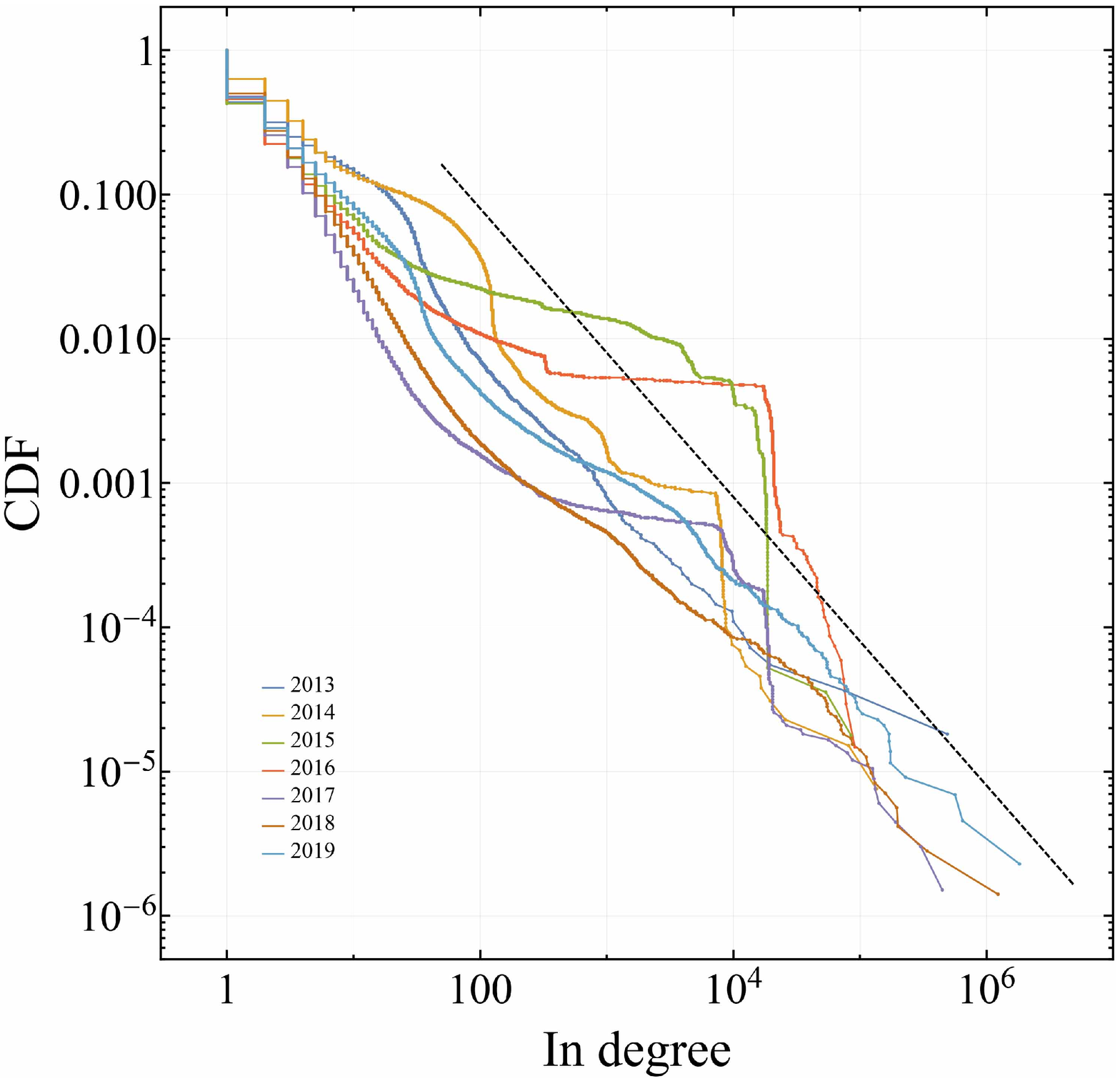}
    \includegraphics[width=0.49\textwidth]{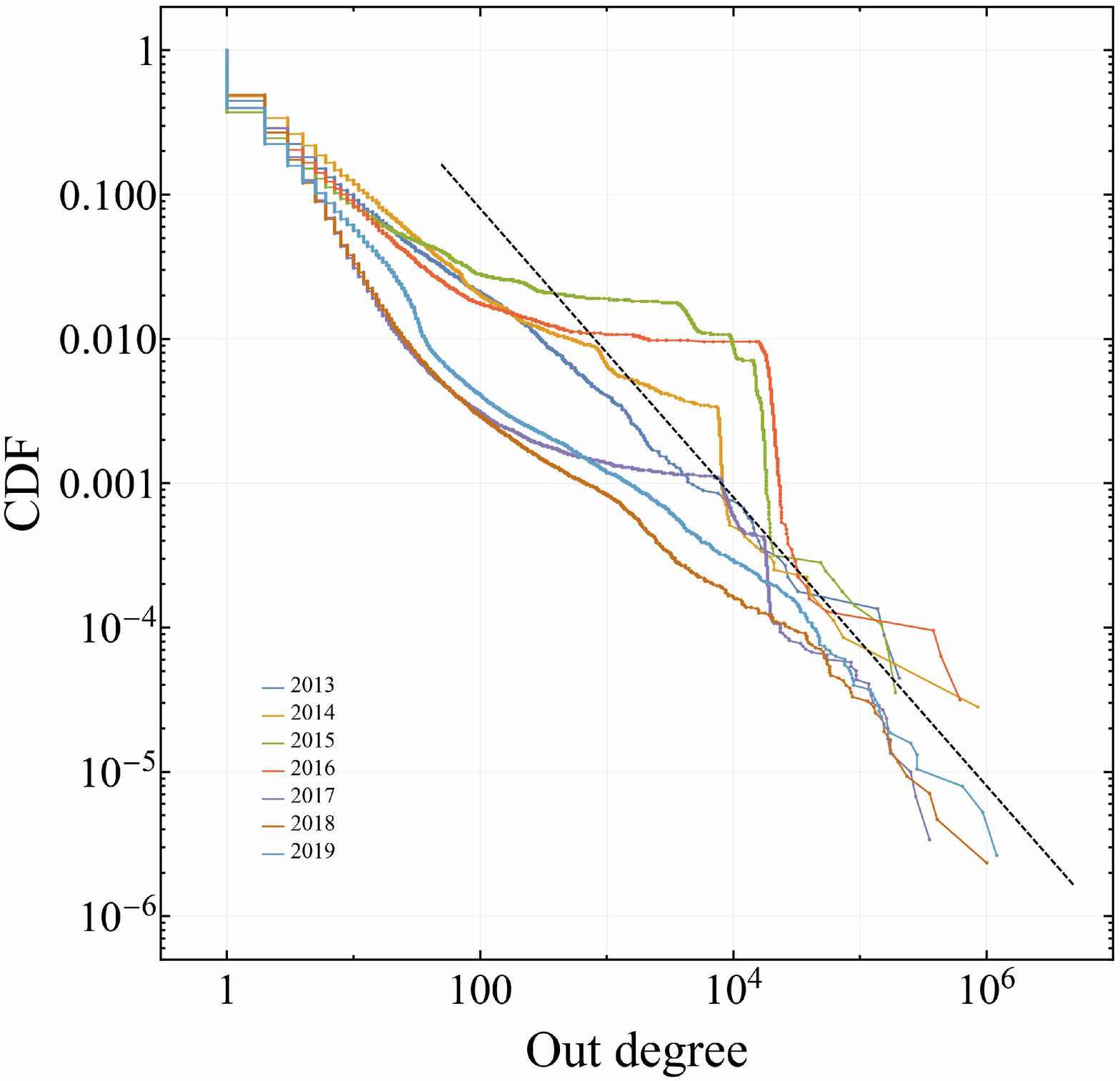}
    \caption{Annual CDF of the in-degree (left) and out-degree (right) of XRP transaction network.}
    \label{fig:inout}
\end{figure}

The CDFs of in-degree and out-degree are plotted in \figref{fig:inout}.
We observe that, except for 2015--2017, it also has a fat tail with Pareto index 1.
Again, this is an interesting finding, awaiting deeper insight and/or modelling.

\subsection{Nodes with Large Transaction History}
\begin{figure}[t] 
    \centering
    \includegraphics[width=0.8\textwidth]{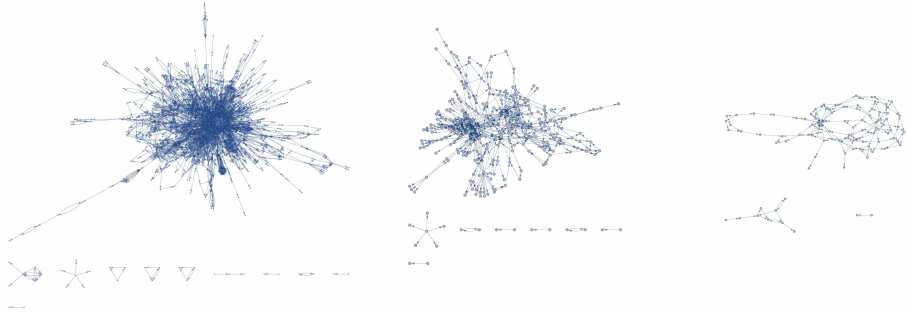}
    \caption{Networks of the nodes selected with three thresholds listed in \tabref{tab:bnsize}.}
    \label{fig:3bnn}
\end{figure}

As noted above, the transactions cover a vast range of $10^{-6}$ (minimum unit) XRP to $\sim 10^{11}$ XRP (\figref{fig:XRPCDF}).
As dealing with them all is unproductive, we introduce a threshold for their biggest transaction.
Let us first  examine nodes with transactions equal to or greater than 
$10^{7,8,9}$ XRP at least once during the entire period (2013--2019).
The network sizes they form are listed in \tabref{tab:bnsize}, and the corresponding networks are visualized in \figref{fig:3bnn}.
In denoting the node, we use the set of 1,136 nodes in the $\ge 10^7$ category and name the nodes with ``bn'' plus its number (0001--1136) in the set. 
(Hereafter, we shall call those 1,136 nodes as ``big nodes,'' and the 94 nodes with threshold $=10^9$ XRP as ``huge nodes.'') 
For example, ``bn0001'' is the first node in a set of big nodes.

\begin{table}[b]
    \centering
    \begin{tabular}{c|rr}
        Threshold & \# Nodes &\# edges \\
         \hline
        $10^7$ & 1,136 & 5,187\\
        $10^8$ &262 & 685\\
        $10^9$ &94 &170 
    \end{tabular}
    \caption{The sizes of the networks formed by nodes with threshold $10^{7,8,9}$ XRP. }
    \label{tab:bnsize}
\end{table}

\subsection{Some Notable Big Nodes}
Some of the nodes that made these huge transactions have a rather notable transaction history, some of which are listed below.

\begin{description}

\item[Pair Nodes]
This is a pair of nodes, among which a large amount of XRP was transferred, with no other notable activity. An example of this type of node is (bn0347, bn0864) at the top two in \tabref{tab:toptra}. Within a minute, $2\times10^{11}$ XRP was transferred from the former to the latter. The former had no other activity, while the latter had numerous transactions considered to be negligible amounts compared to these transactions.
Their transaction histories are provided in \figref{fig:pairs},
where blue dots present the day and amount of transactions as destination (receival of XRP), red dots show those of transactions as source, and green lines show balance, assuming it is zero initially.

\begin{figure}[t]
    \centering
    \includegraphics[width=0.49\textwidth]{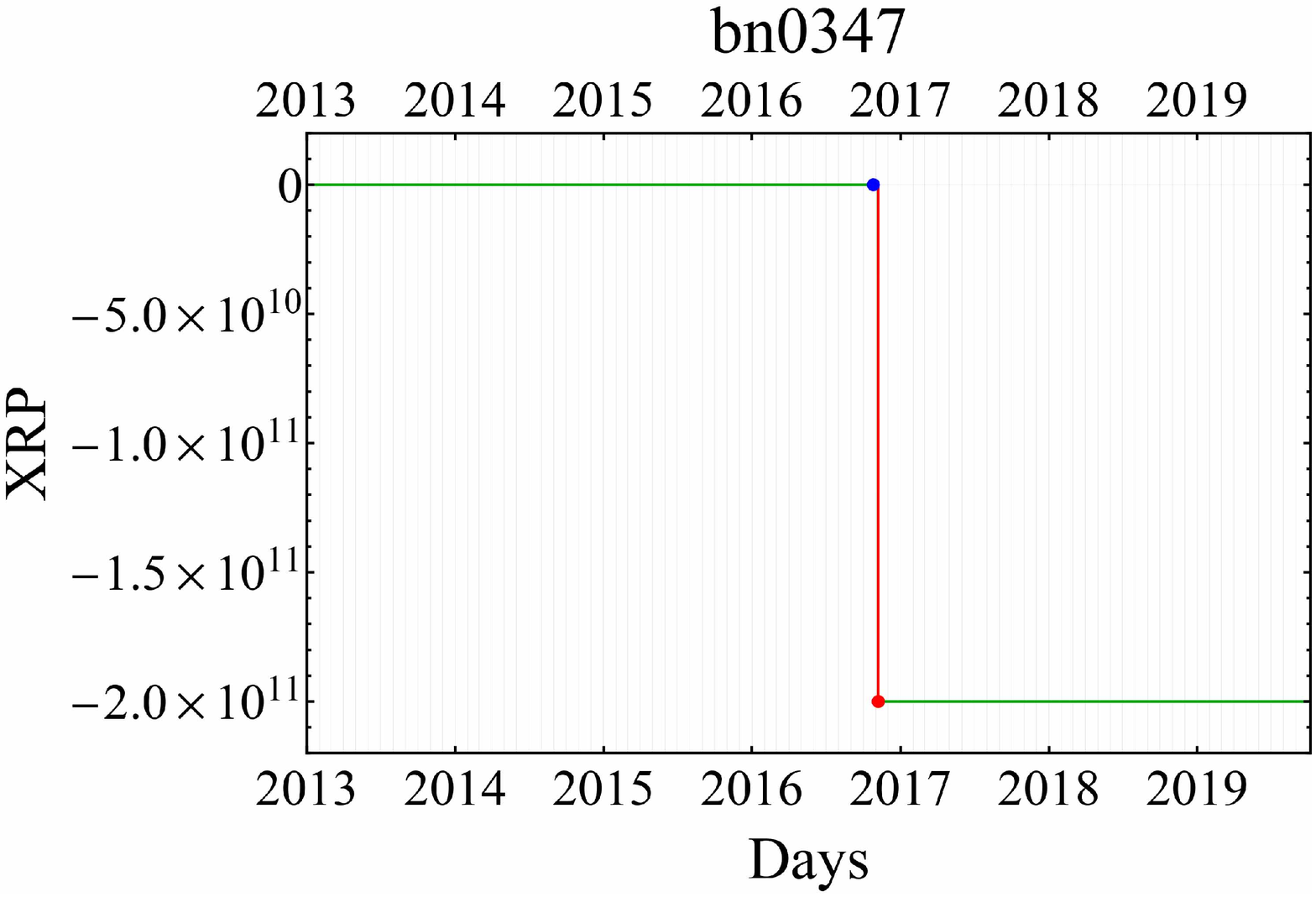}
    \includegraphics[width=0.49\textwidth]{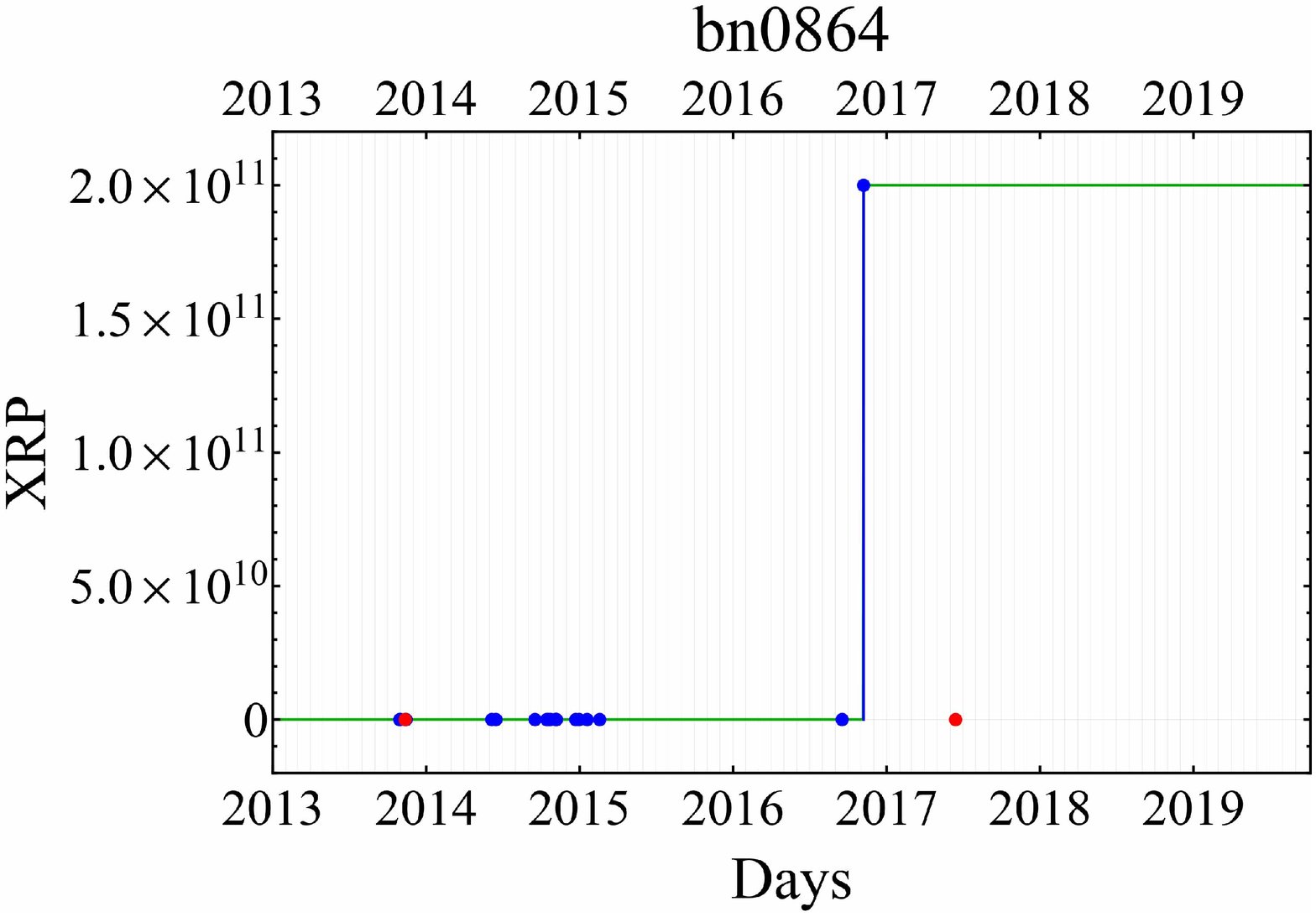}
    \caption{An example of  Pair Nodes.}
    \label{fig:pairs}
\end{figure}

\item[Bridge Nodes]
This node receives a large amount of XRP and sends it to another node,  with no other notable activity.
The node bn530, the third in \tabref{tab:toptra}, is an example of this case, whose transaction history is plotted in \figref{fig:bridge}. 

\begin{figure}[t]
    \centering
    \includegraphics[width=0.7\textwidth]{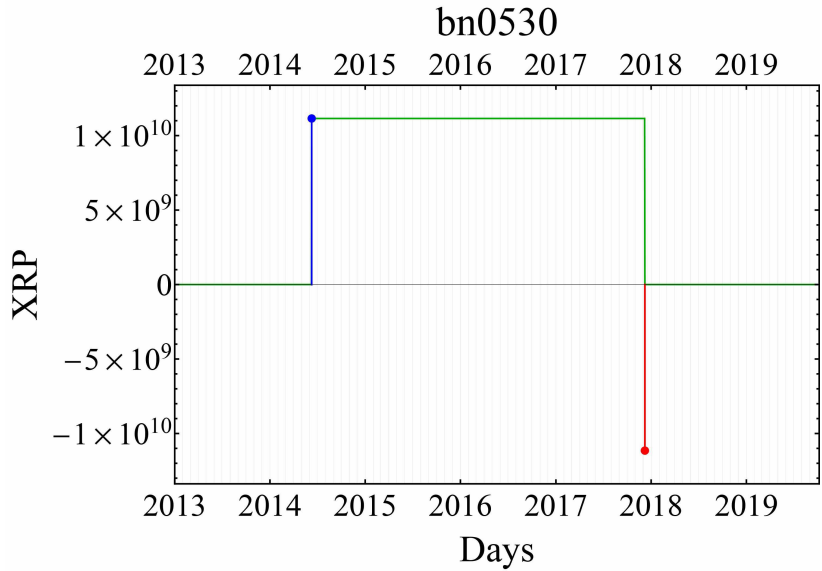}
    \caption{A bridge node.}
    \label{fig:bridge}
\end{figure}
\end{description}

In characterizing these nodes considering the amount and frequency of transactions, noting that some nodes make these huge transactions and many transactions of small amounts is important.
A good example is the node bn0846 in \figref{fig:pairs}:
This node made several transactions of small amounts as both a destination and source. However, they are negligible compared to the two large transactions on the same day, totaling 2.0$\times 10^{11}$ XRP, which are the only significant transactions in characterizing the transaction behavior of this node.
As made clear from this example, a simple count of the number of transactions (as a source or destination) or total number of transactions cannot be considered a good measure of its activity. What counts is the number of ``significant'' (in the amount) transactions it made.
For this purpose, we propose a new index called the ``Flow Index,'' which gives effective number of transactions in each of the directions.

\section{Flow Index}
\subsection{Herfindahl-Hirschman Index}
The Herfindahl-Hirschman Index \cite{rhoades1993herfindahl} (hereafter abbreviated to ``HH Index,''") is used in several data analysis areas to quantify how numbers are distributed to  components in a list.
Consider a list $\ell$ of $N$ non-negative numbers, whose total number is equal to 1:
\begin{equation}
    \ell=\bigl(\ell_1, \ell_2, \cdots, \ell_N \bigr), 
    \quad \ell_m\ge0,
    \quad \sum_{m=1}^N\ell_m =1.
    \label{eqn:ldef1}
\end{equation}
(One might think of this list as a list of shares:
For example, the first entry $\ell_1$ is the share of the 1st firm in the sales of a certain good, $\ell_2$ share of the second firm, and so on.)
Its HH Index $H(\ell)$ is defined as follows:
\begin{equation}
    H(\ell)=\sum_{k=1}^N {\ell_k}^2,
\end{equation}
and satisfies $0< H(\ell)\le1$. 

Here are some examples:
\begin{align}
    \ell=\bigl(1, 0, \cdots, 0 \bigr); & \quad  H(\ell)=1,\\
    \ell=\frac1m\bigl(1, 1, \cdots, \ell_m=1, 0, \cdots, 0 \bigr);  & \quad H(\ell)=1/m.
    \label{eqn:latter}
\end{align}
As presented here, the HH Index $H(\ell)$ is a measure of the concentration of the values in $\ell$:
If it is concentrated to just one component, $H(\ell)=1$.
If it is less concentrated, the smaller $H(\ell)$ is.

The inverse HH Index, $1/H(\ell)$, may be used as a measure of the effective number of entries, as $1/H(\ell)=m$ in the latter case \eqref{eqn:latter}.
However, it has one undesirable property. In the next subsection, we describe the method and propose a modification for overcoming it.

\subsection{Modified \ihh}
\begin{figure}
    \centering
    \includegraphics[width=0.6\textwidth]{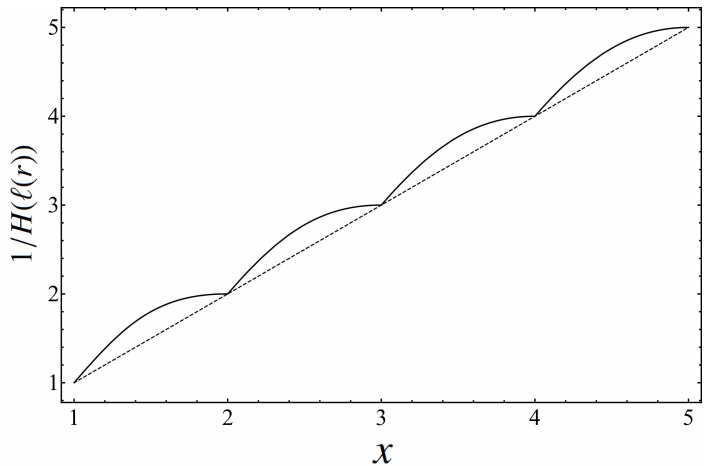}
    \caption{Behavior of the inverse HH Index $1/H(\ell(r))$ as a function of $x=m+r$.}
    \label{fig:invhh1}
\end{figure}

Let us examine the behavior of the inverse HH Index for a generalization of \eqref{eqn:latter}.
\begin{align}
        \ell(r)=\frac1{m+r}\,\bigl( 1, 1, \cdots, \ell_m=1, \ell_{m+1}=r, 0, \cdots, 0 \bigr),
        \label{eq:ellr}
\end{align}
with $0\le r \le 1$,
which has
\begin{equation}
    H(\ell(r))=\frac{m+r^2}{(m+r)^2}.
    \label{eq:hh}
\end{equation}
The inverse is plotted in \figref{fig:invhh1} as a function of $m+r\ (\equiv x)$. As shown here,$1/H(\ell(r))=m$ at integer values of $x=m$ ($r=0$),  as noted above.
However, it flattens as $r\rightarrow 1$, and the derivative with respect to $x$ is discontinuous at $x=m$. 
Essentially, the inverse HH Index is much less sensitive to the reduction in the distribution of numbers ($r$ decreases starting at $r=1$ for fixed $m-1$)
than its expansion ($r$ increases starting at $r=0$ for a fixed $m$).
It also deviates in certain ways from the dashed diagonal line $f(x)=x$, while having a measure closer to this line is desirable.

To overcome this difficulty, we define the following measure, modified \ihh\ of the $n$-th order:
\begin{equation}
    M_n(\ell)=\frac{\bar{H}_{n-1}(\ell)}{\bar{H}_n(\ell)},
    \label{eqn:mhh1}
\end{equation}    
where $\bar{H}_n(\ell)$ is a modified HH Index:
\begin{equation}    
    \bar{H}_n(\ell)=\sum_{k=1}^N \ell_k^n,
    \label{eqn:mhh2}
\end{equation}
identical to HH Index for $n=2$, $\bar{H}_2(\ell)=H(\ell)$.
As $\bar{H}_1(\ell)=1$, $M_2(\ell)$ is the inverse HH Index, $M_2(\ell)=1/H(\ell)$.

\begin{figure}
    \centering
    \includegraphics[width=0.47\textwidth]{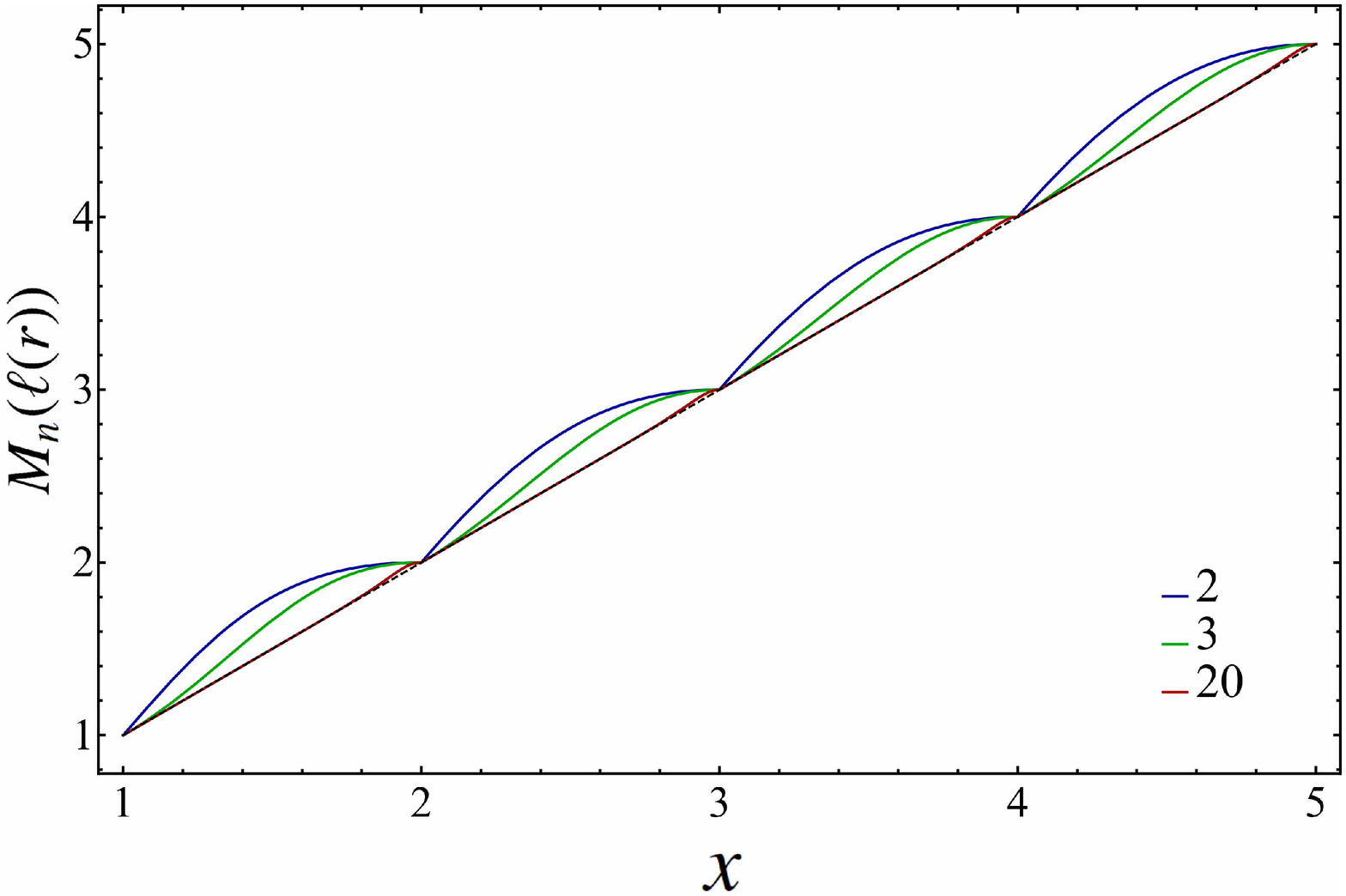}
    \includegraphics[width=0.49\textwidth]{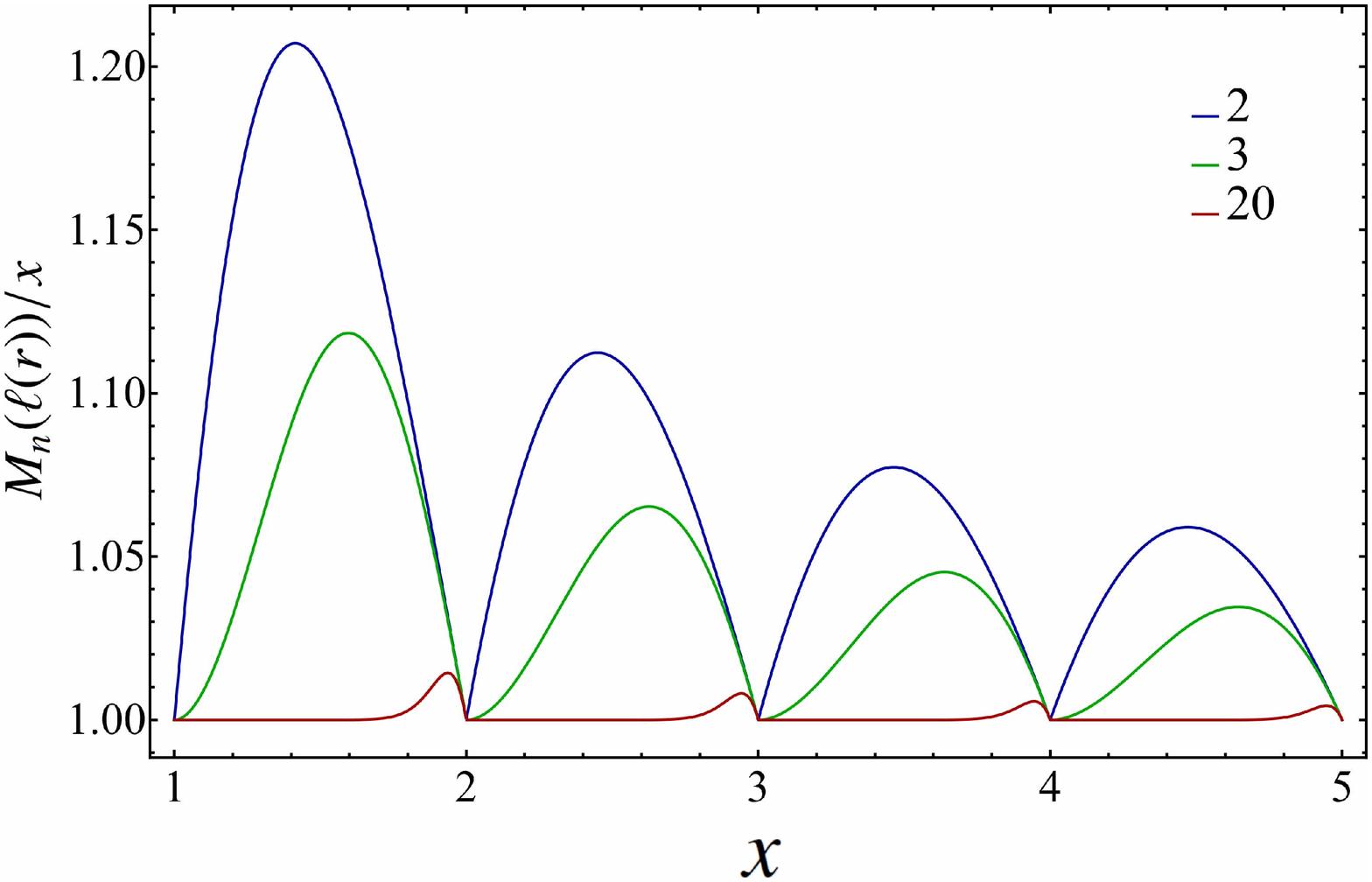}
    \caption{Left: Modified inverse HH Index $M_n(\ell(r))$ as a function of $x=m+r$ for $n=2,3,20$ from top (blue) to bottom (green).
    Right: $M_n(\ell(r))/x$.}
    \label{fig:invhh2}
\end{figure}

For the case of the list $\ell(r)$ in \eqref{eq:ellr},
\begin{equation}
    M_n(\ell(r))=\frac{(m+r)(m+r^{n-1})}{m+r^n},
\end{equation}
and it satisfies $M_n(\ell(0))=m$.
Its behavior for $n=2,3,10$ is plotted in \figref{fig:invhh2}.
The higher $n$, the closer $M_n(\ell(r))$ to $m+r$, 
because $r^n \rightarrow 0$ for $n\rightarrow \infty$ ($0\le r <1$).
\begin{equation}
    \lim_{n\rightarrow \infty}\frac{(m+r)(m+r^{n-1})}{m+r^n}
    \rightarrow m+r.
\end{equation}
Because of this property, one may choose $n$ to be a large number for analysis. In the following, we use $n=20$ because the difference observed in the right panel of \figref{fig:invhh2} is at most $\sim 1.4$\%.

\subsection{Flow Index}
\newcommand\fin{f_{\rm in}}
\newcommand\fo{f_{\rm out}}
\newcommand\bfin{\bar{f}_{\rm in}}
\newcommand\bfo{\bar{f}_{\rm out}}
\newcommand\Max{{\rm Max}}

The modified inverse HH Index defined above is useful for quantifying a node's transaction history.
Let us denote the time series of the daily outflow by $\fo$ and the daily inflow by $\fin$ for this discussion. All the components of $\fo$ and $\fin$ are positive.
Their ``normalized'' (in the sense that the total of all components is  equal to 1, as in Eq.\eqref{eqn:ldef1}) versions are denoted $\bfo$ and $\bfin$.
In a case with no flow, for example, $\fo=\{\}$ (an empty set), we define $\bfo=\{\}$ and $M_n(\bfo)=0$, and so on.

Here we are dealing with the flows aggregated daily.
Alternatively, one may deal with tick data of the flows.
The difference is that if a node makes several large transactions within a short period of time, treating them as one transaction is most appropriate. Daily aggregation takes care of them unless several transactions are made in a time window that includes 0:00 UTC.
For this reason, the daily aggregation is chosen in this study.

Take the node bn0864 shown in the right panel of \figref{fig:pairs}.
This node fits the case discussed above: two transactions of $1.0\times 10^{11}$ XRP each within 50 s of time were made.
Daily aggregation treats them as one $2.0\times 10^{11}$ XRP transaction.
It also included lots of small income over 17 days.  
Therefore, its effective inflow history is best summarized to be on ``(very close to) just one occasion.'' 
Its modified inverse HH Index is, in fact, $M_{20}(\bfin)=1.00005$, quantifying this fact.

This is not the end of the story.
This node had payment over two days (two red dots in the right panel of \figref{fig:pairs}) and its modified inverse HH Index for the outflow is $M_{20}(\bfo)=1.20427$.
However, the amount of outflow was negligible compared to the inflow.
We need to discount the outflow relative to the inflow for this node.

To do so for all nodes, we introduce the following quantity, the {\bf Flow Index}:
\begin{equation}
    A=\left(
    M_{20}(\bfin)\ \frac{\Max(\fin)}{\Max (\fin, \fo)},
    M_{20}(\bfo)\ \frac{\Max(\fo)}{\Max (\fin, \fo)}
    \right).
\end{equation}
$\Max (\fin)$ is the maximum of the components of the outflow, and
$\Max (\fin,\fo)$ denotes the maximum of the components of the joined set of inflows and outflows.

For node bn0864,
we obtain
\begin{equation}
    A=(1.0005, 5.8955\times 10^{-10}),
\end{equation}
a very satisfactory result.

One may think of modifying the above by using the total volume of flows instead of maximum values. However, that does not work. This can be further explored by our readers.

\section{Global Structure of the XRP Network}

\begin{figure}
    \centering
    \includegraphics[width=0.40\textwidth]{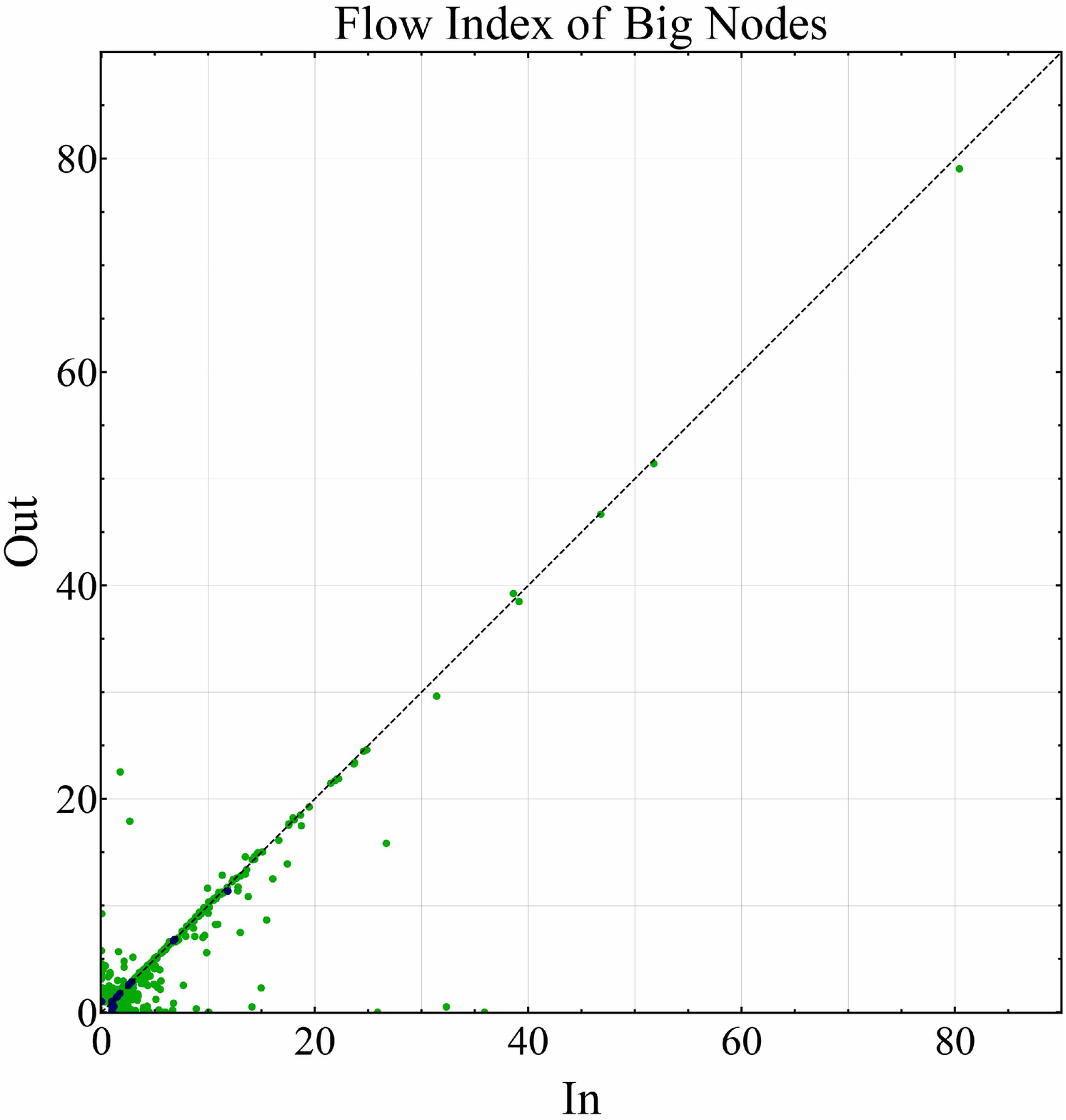}
    \includegraphics[width=0.40\textwidth]{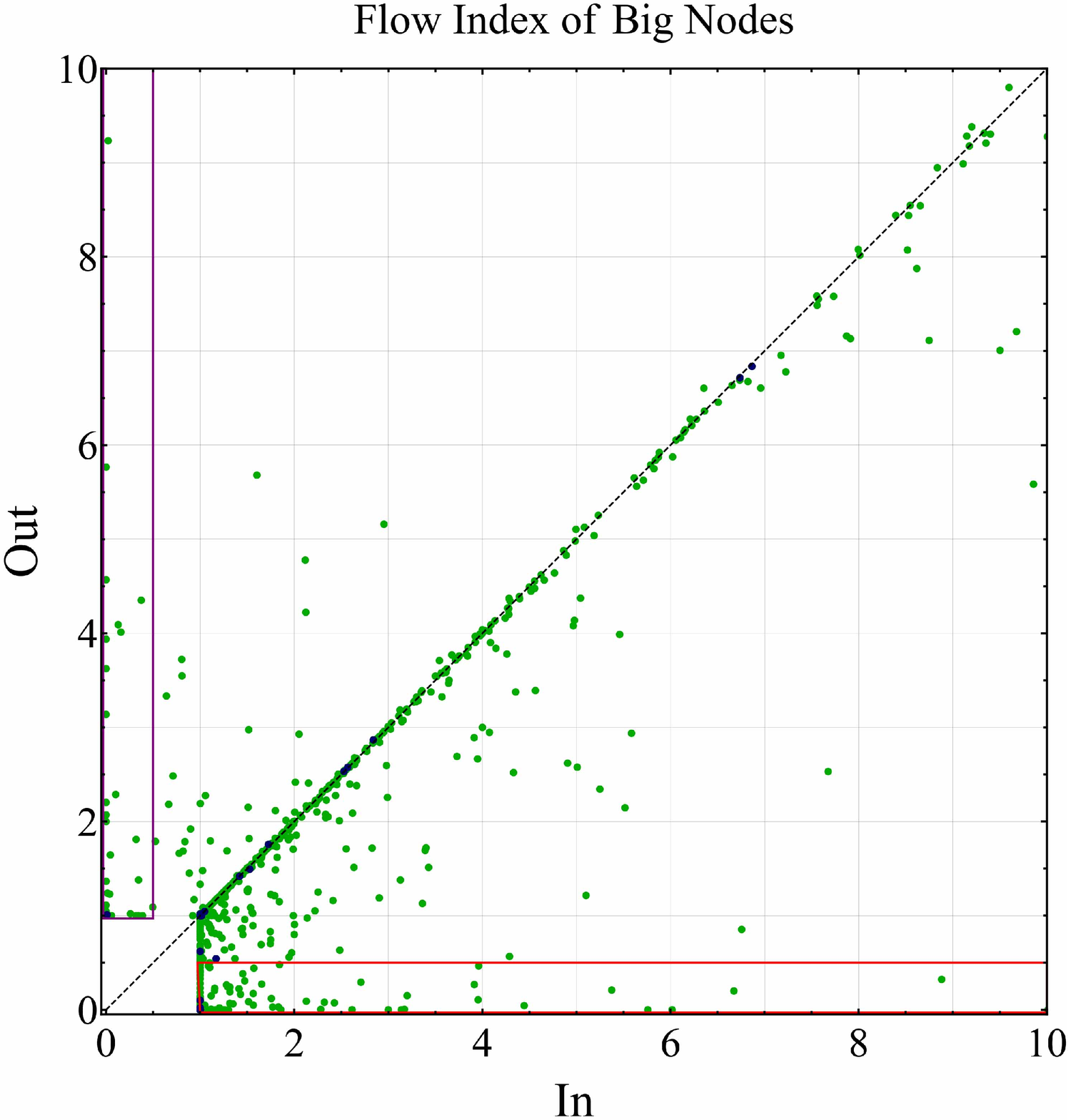}
    \caption{Scatter plot of all big nodes (green) and huge nodes (blue) on the Flow Index plane. The left panel shows all of the big nodes, while the right panel shows details close to the origin.}
    \label{fig:scatterfi}
\end{figure}

The left panel of \figref{fig:scatterfi}
is a scatter plot of 1,136 big nodes (green, threshold $=10^7$ XRP) and 1,176 huge nodes (blue, above threshold $=10^9$ XRP) (see \tabref{tab:bnsize}) on the Flow Index plane. 
We observe that the nodes are distributed somewhat widely on the lower-left part of the Flow Index plane. This implies a tendency of nodes with a small number of effective transactions (as counted by the Flow Index) tend to trade as  destination and as origin unevenly.
In contrast, those with a higher number of transactions are located close to the diagonal, meaning that they tend to trade as destinations and origins evenly. This tendency is true for both big nodes and huge nodes.

The right panel of \figref{fig:scatterfi} shows the details of the left panel. 
We observe the existence of nodes that mostly participate in one mode. Motivated by this type of distribution, we classify  the nodes with $A_1\le0.5$ (red rectangle) as ``OUT'' components,
as they are mostly on the destination side. This means they are at the final goal of XRP when viewed as part of the whole network. We found 193 nodes.
Similarly, we classify the nodes with $A_2\le 0.5$ (purple rectangle) as ``IN'' components; there are 52 of them.

\begin{figure}
    \centering
    \includegraphics[width=0.7\textwidth]{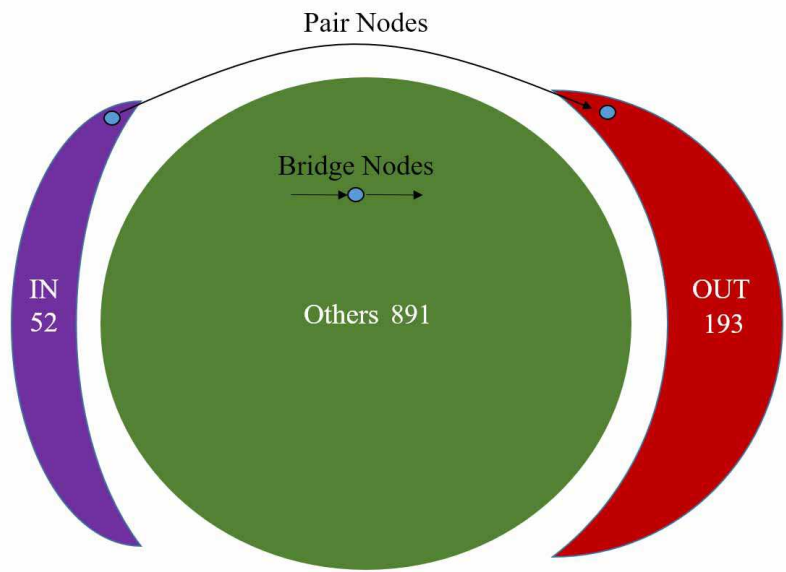}
    \caption{The bow-tie/walnut-like structure of big nodes obtained by the use of Flow Index.}
    \label{fig:walnut}
\end{figure}

Using this criteria, we can draw
 bow-tie/walnut-like structures \cite{chakraborty2018hierarchical}, as shown in \figref{fig:walnut}.

This characterizes the global structure of the XRP network, which forms the basis for understanding the dynamics and development of this complex structure. 

\clearpage
\section{Concluding Remarks}

In this study, we presented an analysis of XRP transaction records from ledger data.

These data are huge and complex: 
In addition to the number of transactions,
The distribution of traded amount, frequency of transactions, and so on cover huge ranges, some of which cover 18 orders of magnitude ($10^{-6}$ to $10^{11}$ XRP). 
A notable empirical finding includes the power distribution of several quantities with a Pareto exponent close to one, and a power-law correlation between the daily number of transactions and the daily amount of transactions.
The former remains a puzzle: While Pareto index equal to one is known to be the beginning of the monopoly/oligopoly phase, we have a reasonable explanation behind it only for stock quantities like
 number of employees at firms. The current transaction amount between nodes is a flow..
The latter, the power-law correlation, may be explained in terms of ``herding behavior.'' 
Proper modelling with the use of deeper analysis of the current data should lead to the explanation of this correlation. These subjects are worth more extensive exploration in the future.

The XRP network, a directive network with nodes as transaction accounts and edges as transactions, is another main subject of this research.
To concentrate on the central structure of this network, we placed a threshold for the maximum amount of each node. We selected nodes that made transactions of more than $\ge 10^7$ XRP at least once and called them big nodes.
To examine transaction frequency while considering the huge range of transaction amount each nodes make, we defined a new index called ``Flow Index,'' borrowing and extending the idea of the Herfindahl-Hirschman Index.
We further introduced classification of nodes using the Flow Index and arrived at a view of the entire network as a  bowtie/walnut-like.

We believe this work establishes a foundation for not only the XRP network but also other dynamic networks of transactions.
Further research on this network should reveal details of the activities and clarification of each node's characteristics on the structure of the bow-tie/walnut-like decompositions.

\noindent{\sl Note added in proof:}

Toward the end of writing this manuscript, the author learned of a new paper by Fujiwara and Islam  \cite{YoshiNow}, where they examined the Bitcoin network formed by ``regular users.'' This approach is complementary to our current analysis. While the latter chooses to select users based on the number of transactions, the former analyzes the frequency of transactions.
The current author believes that a new approach based on both ways of thinking and picking up good features from both is waiting for us in the near future.

\section*{Acknowledgments}
The author would like to acknowledge Ripple, which is providing financial and technical support through its University Blockchain Research Initiative.
He would also like to thank 
Yuichi Ikeda for providing him with the ledger data, 
Yoshi Fujiwara and Hiro Inoue for technical support on data-handling, 
Yuzuki Nomura for helping him with text entry and editing,
and Editage (www.editage.com) for English language editing.


\end{document}